\documentclass[review]{elsarticle}

\usepackage{lineno,hyperref}
\usepackage{color}
\usepackage{amssymb}
\usepackage{graphics}
\usepackage{graphicx}
\usepackage{pbox}
\usepackage{mhchem}

\usepackage{xcolor} %colors,yay
\usepackage[ampersand]{easylist}
\usepackage{lineno}
\usepackage{booktabs}
\usepackage{multirow}
\usepackage{subcaption}
\usepackage{amsmath}
\usepackage{paralist}
\usepackage{soul}
\usepackage{siunitx}
\usepackage{amsmath}
\usepackage{url}
\usepackage{makecell}
\usepackage{xcolor}
\usepackage{lineno}

\biboptions{sort&compress}

\hypersetup{  %UNCOMENT
    colorlinks=true,
    linkcolor=blue,
    filecolor=magenta,      
    urlcolor=cyan,
    pdftitle={Sharelatex Example},
    bookmarks=true,
    pdfpagemode=FullScreen,
    citecolor=blue
    }

\date{}

\journal{NIM A}

\begin{document}

\begin{frontmatter}

\title{Characterization of GS20 and CLYC Detectors for Neutron Resonance Transmission Analysis in High Radiation Environments}

\author[MITaddress]{Shayaan Subzwari}
\author[PNNL]{Benjamin McDonald}
\author[MITaddress]{Areg Danagoulian\corref{corauthor}}
\cortext[corauthor]{Corresponding author}
\ead{aregjan@mit.edu}

\address[MITaddress]{Massachusetts Institute of Technology, Cambridge, MA 02139, USA}
\address[PNNL]{Pacific Northwest National Laboratory, Washington, USA}

\begin{abstract}
Advanced reactor concepts based on the thorium fuel cycle offer several advantages over conventional uranium-fueled systems, but they also stress-test the existing non-destructive assay (NDA) toolbox for international safeguards. In particular, the presence of \textsuperscript{232}U and its $\sim$MeV gamma-emitting daughters in thorium-based spent fuel creates a harsh radiological environment that complicates gamma-based active interrogation safeguard techniques. Neutron resonance transmission analysis (NRTA) has emerged as a promising safeguards technique due to its isotopic specificity in the epithermal range and its robustness against non-resonant shielding. However, deploying NRTA in thorium safeguards requires neutron detectors that maintain timing performance and quantitative accuracy in intense gamma fields. 

This paper reports a comparative characterization of two candidate detectors for portable NRTA: \textsuperscript{6}Li:Ce glass (GS20) and Cs\textsubscript{2}LiYCl\textsubscript{6}:Ce (CLYC). GS20 has already been demonstrated as an effective epithermal detector in portable NRTA systems but offers limited neutron--gamma discrimination. CLYC, by contrast, provides strong pulse-shape discrimination (PSD) but has a much longer scintillation decay time and includes \textsuperscript{133}Cs, whose resonances partially overlap with key actinide resonances in the epithermal region. 

Using a D--T--driven NRTA setup with a 2~m flight path, we compare GS20 and CLYC in measurements of a 1.50~mm tungsten target under both ``clean'' conditions and in an artificially constructed high $\gamma$-radiation environment produced by an auxiliary source as a way of emulating a highly radioactive \textsuperscript{233}U target. Transmission spectra are analyzed with the NeuFIT open-source fitting code, which incorporates setup-specific time-of-flight response functions to generate physically realistic model spectra and to extract the target areal densities~\cite{Subzwari_NRTA_Thesis_2025}. Both detectors recover the tungsten thickness with good accuracy in all conditions, but the associated uncertainties behave very differently: CLYC yields consistently smaller errors, and unlike GS20 its performance is essentially unchanged when the gamma background is increased to levels comparable to previous \textsuperscript{233}U measurements. 

These results indicate that CLYC, despite its long decay time and intrinsic \textsuperscript{133}Cs resonances, provides significantly more precise NRTA measurements in high radiation environments than GS20. For thorium-based safeguards scenarios where \textsuperscript{233}U must be identified and quantified in the presence of intense gamma backgrounds, CLYC-like detectors with strong PSD appear to be the more reliable choice, provided that the internal resonances do not obscure the target absorption lines.
\end{abstract}

\end{frontmatter}

\section{Introduction}
%\linenumbers
Nuclear energy deployment has always been tightly coupled to questions of nonproliferation, verification, and safeguards. Under the Treaty on the Non-Proliferation of Nuclear Weapons (NPT), non-nuclear-weapon states are obligated to accept International Atomic Energy Agency (IAEA) safeguards intended to detect diversion of nuclear material from peaceful uses to unsanctioned purposes. In practice, this means credibly quantifying and tracking special fissionable material---such as plutonium and enriched uranium---as well as fertile materials like depleted uranium and thorium throughout the fuel cycle.

For conventional light-water reactors based on the \textsuperscript{235}U/\textsuperscript{238}U fuel cycle, the safeguards toolkit is mature. Gamma-ray spectroscopy, passive and active neutron counting, and well-calibrated destructive assay methods together cover most inspection needs~\cite{geist2024nondestructive,Zendel2011}. These methods exploit well-known signatures such as the 186~keV gamma line of \textsuperscript{235}U, characteristic plutonium gamma lines, and spontaneous fission neutrons from curium in spent fuel. However, the emergence of advanced reactor concepts, particularly those built around the thorium fuel cycle, challenges many of the assumptions that underlie these established methods.

In a thorium cycle, fertile \textsuperscript{232}Th breeds fissile \textsuperscript{233}U through neutron capture and subsequent beta decays. Depending on the fuel design, this may occur in solid oxide fuels, molten salt fuels, or composite fuels containing a mix of \textsuperscript{232}Th, \textsuperscript{233}U, \textsuperscript{235}U, \textsuperscript{238}U, and possibly plutonium. Thorium's high capture cross-section and the neutron economy of \textsuperscript{233}U enable thermal-spectrum breeding, but the same fuel cycle also produces trace \textsuperscript{232}U. The decay chain of \textsuperscript{232}U includes \textsuperscript{208}Tl, a strong 2.6 MeV gamma emitter, making recycled thorium fuel and separated \textsuperscript{233}U highly radioactive in the gamma channel.

This gamma activity is sometimes touted as ``self-protection''---a practical deterrent to diversion and weaponization---but it simultaneously undercuts many existing NDA methods. Intense gamma fields complicate passive gamma spectroscopy, saturate detectors, raise backgrounds in neutron detectors, and force the use of heavy shielding or large standoff distances, all of which eat into statistics and sensitivity. Moreover, traditional active neutron interrogation tools like active well coincidence counters can determine total fissile mass but cannot reliably disentangle \textsuperscript{233}U from \textsuperscript{235}U in mixed samples.  

Oak Ridge National Laboratory has highlighted several specific NDA challenges posed by thorium fuel cycles: quantifying \textsuperscript{232}Th, identifying \textsuperscript{233}U behind significant shielding, discriminating \textsuperscript{233}U from \textsuperscript{235}U, and maintaining detector performance in high-gamma environments created by \textsuperscript{232}U daughters~\cite{ORNL_Th}.   These challenges motivate the exploration of NRTA as a safeguards tool and, critically, the identification of detectors that enable NRTA to function under harsh radiological conditions.

\section{Background}

\subsection{Principles of neutron resonance transmission analysis}

NRTA exploits the fact that many mid- and high-$Z$ nuclei exhibit sharp neutron absorption resonances in the epithermal energy range (\SI{1}{eV}--\SI{100}{eV}) and has been used in a variety of nuclear safeguards and arms control applications~\cite{Klein_thesis,Klein_NRTA_2021,pnnl_report,mcdonald2024neutron,azzoune2025conceptual,engel2019physically}.   When the incident neutron energy matches a compound-nucleus level, the capture cross-section spikes according to the Breit--Wigner formalism, and the total neutron cross-section displays a corresponding peak. For a slab of material with thickness $x$, number density $n=\rho N_{Av}/A$, and energy-dependent total cross-section $\sigma(E)$, the neutron transmission is $T(E)=\exp\left[-n\,\sigma(E)\,x\right]$. Resonances thus appear as dips in the measured transmission spectrum. Because the set of resonance energies and widths is unique to each isotope, the pattern of dips encodes both the isotopic composition and the areal densities present.  

In a typical portable NRTA configuration, a pulsed D--T neutron generator produces \SI{14.1}{MeV} neutrons that are moderated to the epithermal range using a polyethylene-based moderator surrounding a lead multiplier. The moderated flux emerges toward a target positioned downstream; transmitted neutrons are recorded by a fast scintillation detector placed roughly \SI{2}{m} away.   Time-of-flight (TOF) from the neutron pulse to the detection event is used to reconstruct neutron energy:
\begin{equation}
E_n = \frac{1}{2} m_n \left(\frac{\ell_{\text{eff}}}{t_{\text{TOF}}}\right)^2,
\end{equation}
where $\ell_{\text{eff}}$ is an effective path length that accounts for moderation and detector penetration and $t_{\text{TOF}}$ is the corrected TOF between pulse emission and signal detection.  

Transmission is obtained by comparing ``open-beam'' (target-out) and ``target-in'' TOF spectra. Experimental transmission $T_{\text{exp}}(E)$ is then fit with model spectra generated from evaluated nuclear data to determine areal densities for candidate isotopes. NeuFIT, an open-source analysis code, improves on older black-box fitting tools by incorporating setup-specific TOF response functions derived from Geant4 simulations and by performing non-linear least-squares fits via a transparent algorithmic pipeline.  

\section{Detector Requirements and Considerations}

There are three main requirements for a detector suitable for the relevant NRTA applications: the detector must be able to detect epithermal neutrons, it must have a timing resolution better than 1 $\mu$s, and it must have either low gamma sensitivity or effective neutron-gamma discrimination. A range of detectors have been previously considered for the portable NRTA setup, including both scintillators and gas-based detectors. Given the importance of time resolution, scintillators have been largely considered for their relatively faster rise times (1 to 50 ns) compared to gas-based detectors (100 to 500 ns) \cite{pnnl_report}. With high neutron capture cross-sections for thermal and epithermal energy ranges, scintillators with $^6$Li or $^{10}$B are the main focus of consideration.

Across a wide range of plausible $^6$Li detectors available, two of particular interest are $^6$Li:Ce (GS20) glass and Cs$_2$LiYCl$_6$:Ce (CLYC). 

\begin{table}[ht]
\centering
\caption{Comparison of GS20 and CLYC properties \cite{ejik}.}
\label{tab:gs20-clyc}
\begin{tabular}{lcc}
\toprule
Property & GS20 & CLYC \\
\midrule
Density (g/cm$^3$)                                  & 2.50       & 3.31           \\
Absorption length for thermal neutron (mm) & 0.52      & 3.2           \\
Emission wavelength $\lambda_{\mathrm{em}}$ (nm)     & 395       & 380           \\
Light emission per neutron (photons)                 & $\sim7\,000$  & 70\,000       \\
Light emission per MeV gamma (photons)               & $\sim4\,000$  & 20\,000       \\
$\alpha/\beta$ ratio                                 & 0.35      & 0.73          \\
$\Delta$E/E thermal neutron peak                     & 15\%        & 5\% \cite{ferrulli}            \\
Decay time $\tau$ (neutron, ns)                      & 70        & 50 \\
Decay time $\tau$ (gamma, ns)                       & 70        & 50 \\
PHA                                                  & Yes         & Yes             \\
PSD                                                  & No         & Yes             \\
\bottomrule
\end{tabular}
\end{table}

\subsection{$^6$Li:Ce (GS20) glass}

GS20 glass is a popular scintillator material for thermal and epithermal neutron detection, with 6.6\% Li by weight and 95\% isotopic $^6$Li enrichment. Compared to CLYC, the scintillator has a relatively short mean free path -- approximately 0.52 mm for thermal neutrons -- and fast decay time on the order of nanoseconds for both neutrons and gammas \cite{ejik}. While it does have a relatively low light yield and lower energy resolution, its short mean free path and decay time provide it with high thermal detection efficiency and fast pulse timing. 

As considered in previous works, GS20 provides high epithermal neutron detection efficiency, ranging from about 90\% at 1 eV to approximately 10\% at 100 eV \cite{Klein_thesis} for a 5-mm thick scintillator. Increasing the thickness would cause a corresponding increase in neutron efficiencies, although it would concurrently increase gamma detection efficiencies. Given the unwanted contributions of gamma detections to the background signal, previous work has determined that a thickness of 5 mm presents a workable tradeoff between maximizing neutron detection and minimizing gamma backgrounds. 

\begin{figure}[htbp]
    \centering
    \begin{subfigure}[b]{0.49\textwidth}  % Adjust width to your needs
        \centering
        \includegraphics[width=\textwidth]{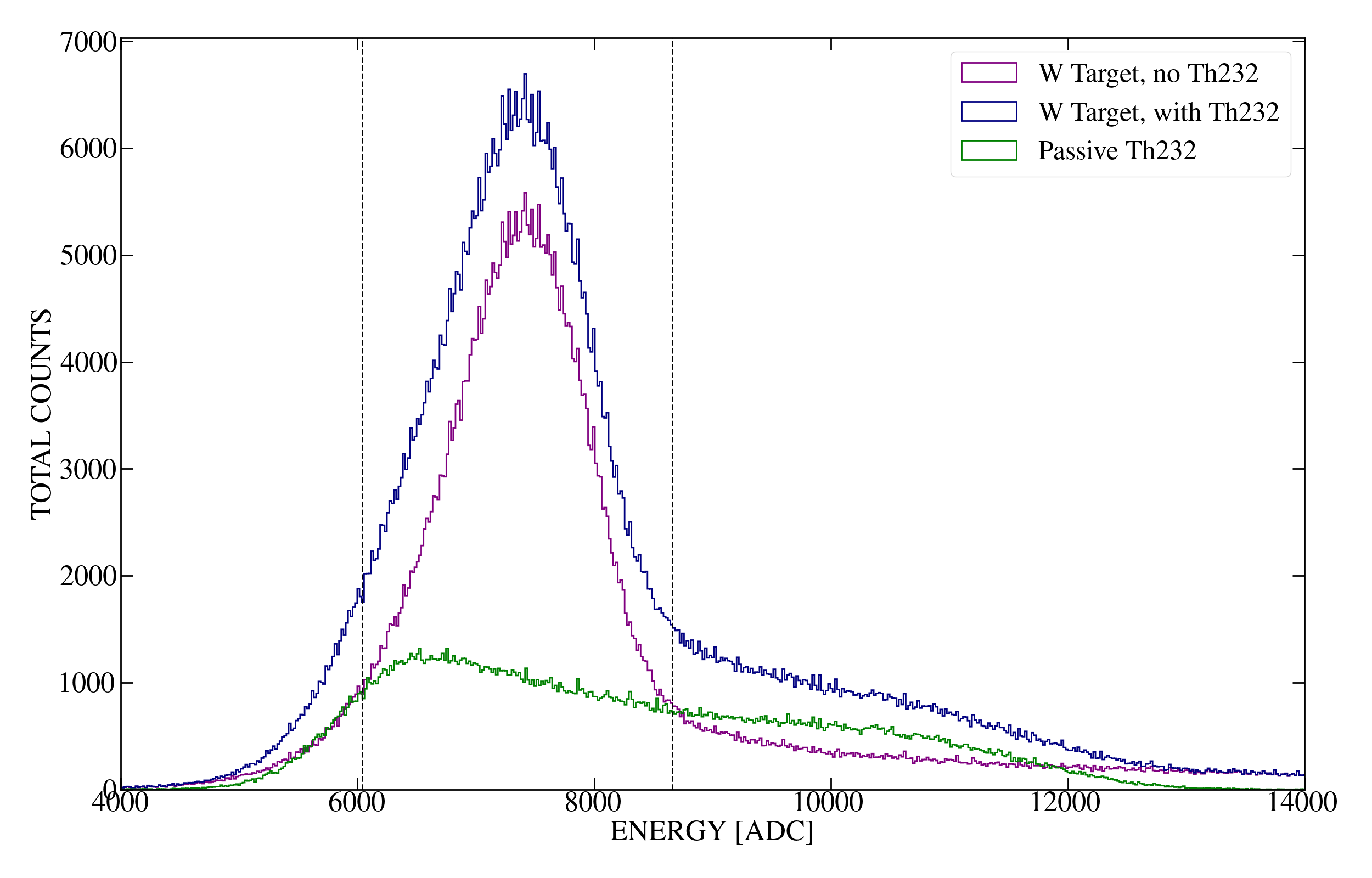}
        \caption{PHA plot from GS20, showing additional gamma contributions when a gamma-emitting target is used. The delineated energy region corresponds to the epithermal filter.}
        \label{fig:PHA}
    \end{subfigure}
    \hfill
    \begin{subfigure}[b]{0.49\textwidth}  % Adjust width to your needs
        \centering
        \includegraphics[width=\textwidth]{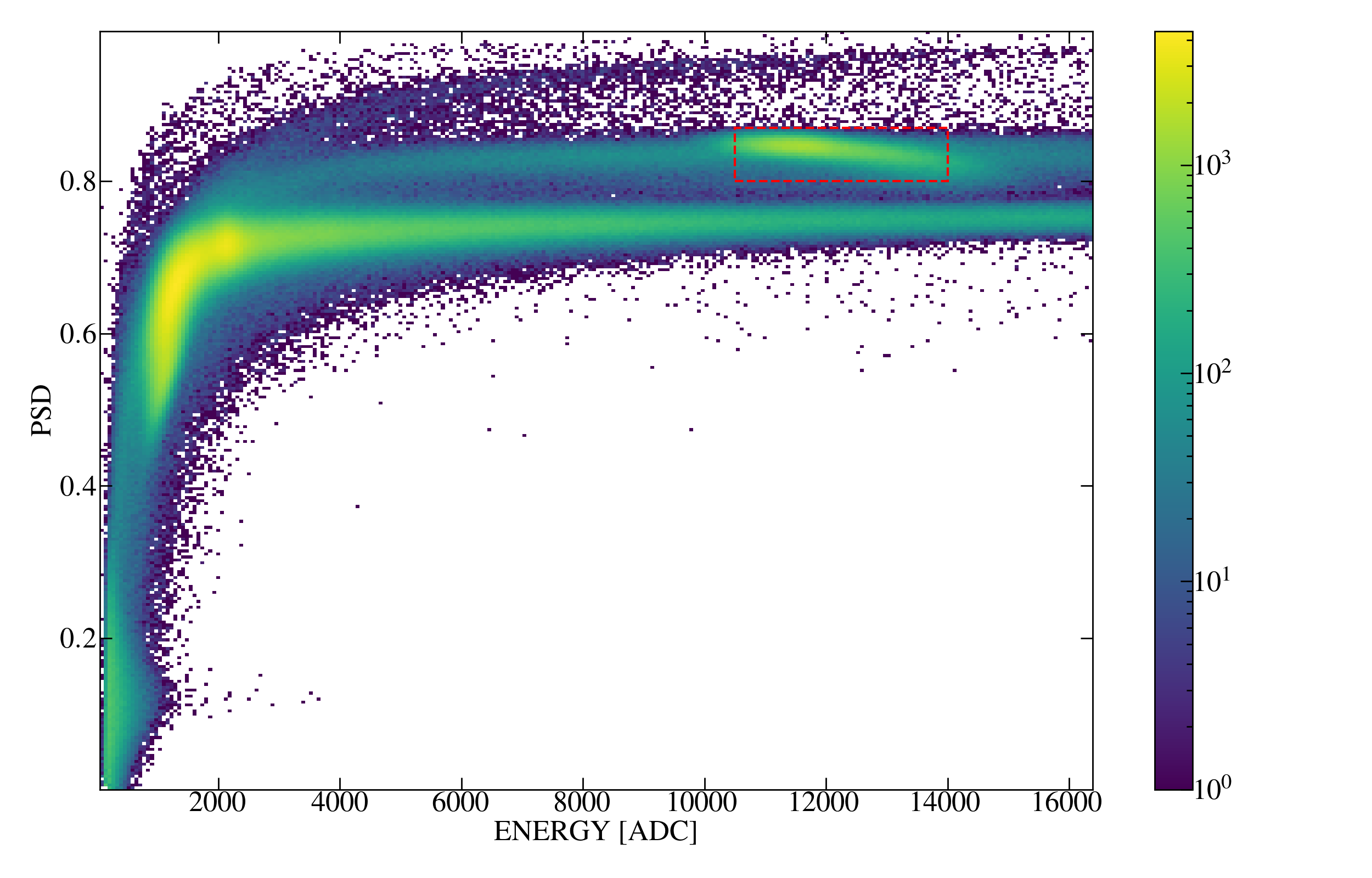}
        \caption{PSD plot from CLYC detector with D-T generator on, with delineated area denoting the epithermal region and lower region corresponding to gammas.}
        \label{fig:clyc_psd}
    \end{subfigure}
    \caption{Neutron-gamma discrimination methods for GS20 and CLYC.}
\end{figure}

Regarding neutron-gamma discrimination, GS20 offers pulse-height analysis (PHA). This is based on the difference in light output from tritons and alphas from detected neutrons and the light output from scattered gammas on the detector. As can be seen in Fig.~\ref{fig:PHA}, the neutron peak is localized while the gamma distribution represents a spread in energy, although substantial gamma counts are still present under the peak corresponding to neutron light output. Performing a cut on this peak allows for a degree of neutron-gamma discrimination, although it retains an estimated 2.5\% gamma background contribution relative to the epithermal neutron signal, as determined by previous analysis \cite{Klein_thesis}. Note that this is solely the contribution from gammas produced in the moderator, in a scenario where there is no effective passive gamma emissions from the target. In the case where there is a significant degree of passive gamma emissions, especially when the corresponding light output can overlap with the peak representing light emission from neutrons, a substantially higher background percentage can result, as seen in Fig.~\ref{fig:PHA}. 

\subsection{Cs$_2$LiYCl$_6$:Ce (CLYC)}

\begin{figure}[htbp]
    \centering
    \begin{subfigure}[b]{0.49\textwidth}  % Adjust width to your needs
        \centering
        \includegraphics[width=\textwidth]{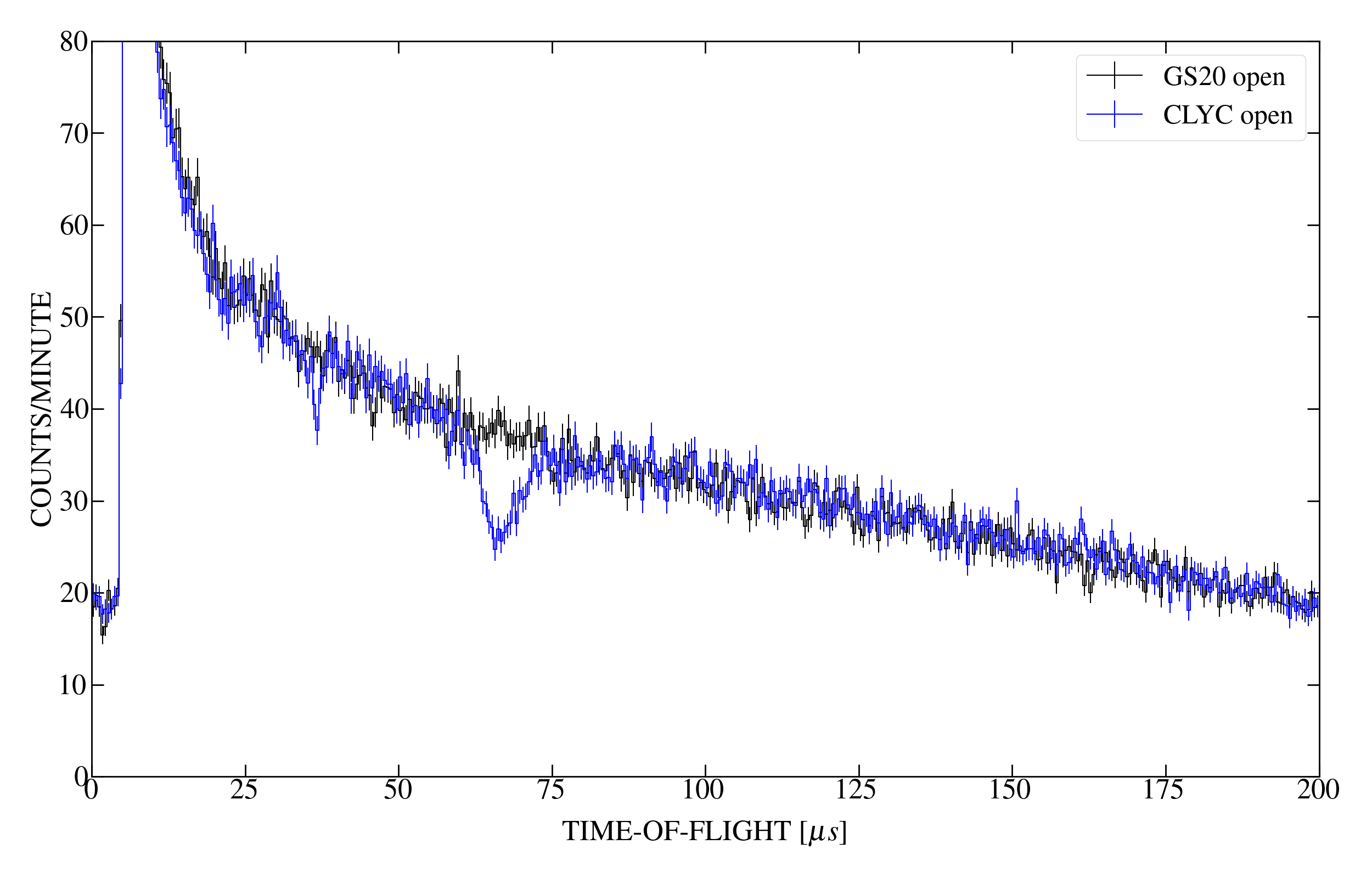}
        \caption{Count rate comparison between selected CLYC and GS20 thicknesses used in experimentation.}
        \label{fig:CLYC_GS20_efficiencies}
    \end{subfigure}
    \hfill
    \begin{subfigure}[b]{0.49\textwidth}  % Adjust width to your needs
        \centering
        \includegraphics[width=0.93\textwidth]{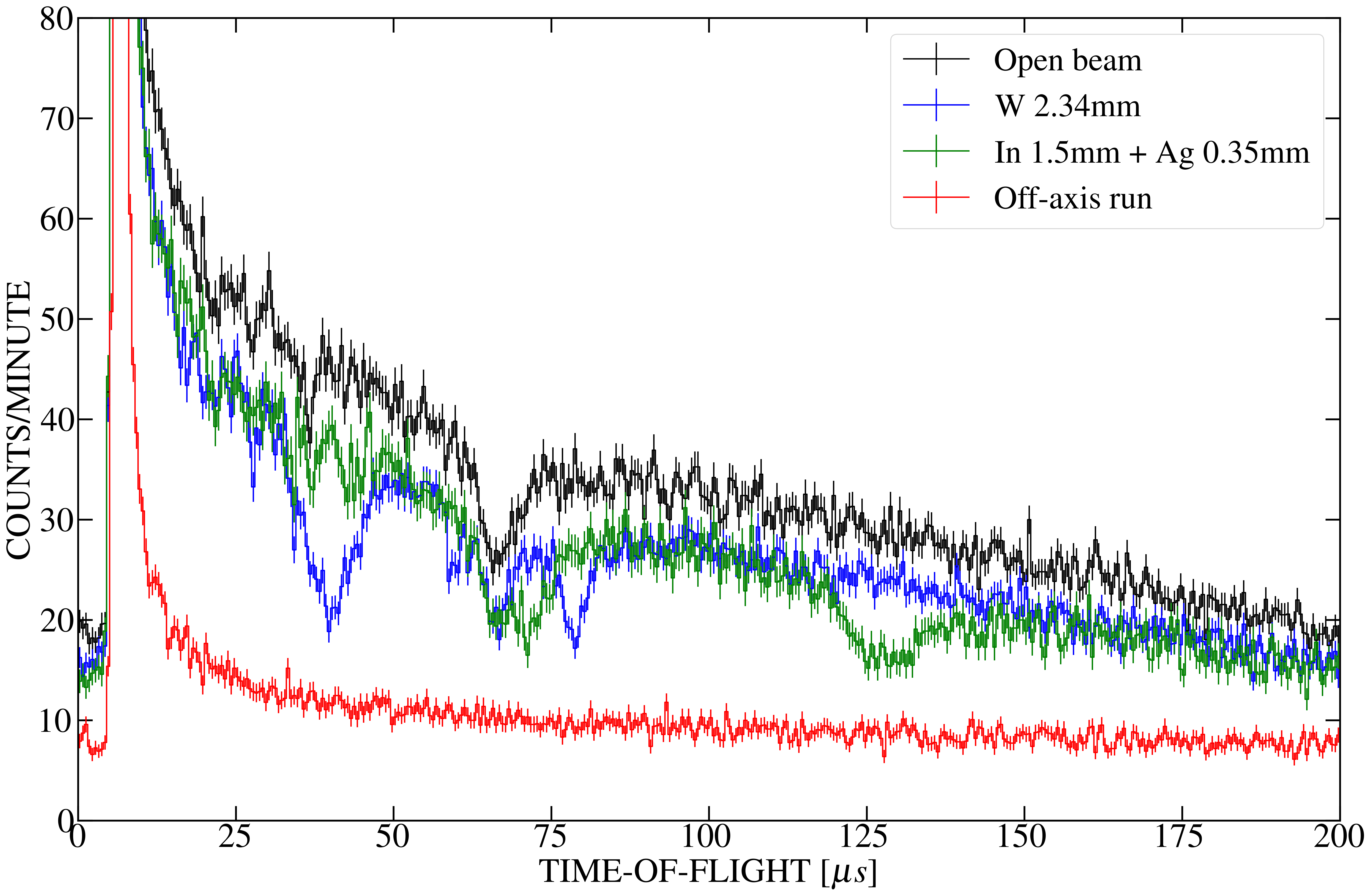}
        \caption{Runs involving various targets for background model construction for CLYC detector.}
        \label{fig:clyc_bg}
    \end{subfigure}
    \caption{Count rates for CLYC and GS20, for efficiency comparison and background model construction.}
\end{figure}

CLYC crystals are another common $^6$Li scintillator with dual neutron-gamma detection capability, containing approximately 95\% $^6$Li enrichment. Relative to GS20, CLYC has a significantly larger absorption length for thermal neutrons, requiring thicker amounts of the crystal to obtain the same neutron efficiency. In order to afford a direct comparison between CLYC and GS20 holding efficiency effects constant, an appropriate thickness of 1 cm of CLYC was chosen in all subsequent analyses in order to yield equivalent epithermal detection efficiencies between both CLYC and GS20, as seen in Fig.~\ref{fig:CLYC_GS20_efficiencies}. 

While it has significantly greater energy resolution and light output compared to GS20, another disadvantage of CLYC is that it is much slower, with a decay time on the scale of microseconds compared to nanoseconds for GS20. This can cause concerns regarding efficiency and pileup in the case of large count rates. In scenarios where there are both incident neutrons and significant passive gamma emissivity, due diligence is needed to ensure that the count rate is not too large to overwhelm the detector and lead to pileup. In the experimental runs detailed in the following sections, an analysis of the detector was conducted to ensure that such situations were not an issue for the expected count rates from the detector and a passive gamma-emitting target. 
If it is indeed found that pileup is not a limiting factor, then the slower timing associated with CLYC is acceptable given the tradeoff that it provides with increased energy resolution, light output, and neutron-gamma discrimination.

One of CLYC's crucial advantages over GS20 is its neutron-gamma discrimination through pulse-shape discrimination (PSD), capturing differences in the waveforms generated by neutrons and gammas depositing energy in the scintillator. As seen in Fig.~\ref{fig:clyc_psd}, a clear delineation can be made between the region associated with gammas and that associated with thermal and epithermal neutrons. Performing a cut in both the energy and PSD directions can allow for an effective filter removing unwanted gamma contributions, compared to PHA, which only allows for a cut in the energy direction. Additionally, given the larger neutron light output for CLYC, the associated peak in the energy direction is also shifted farther to the right, resulting in fewer gammas from the moderator overlapping in energy terms. These factors all lead to an improved rejection of gamma events through PSD relative to PHA. 

\begin{figure}[p]
\captionsetup{justification=centering}
\centering
\includegraphics[width=0.9\textwidth]{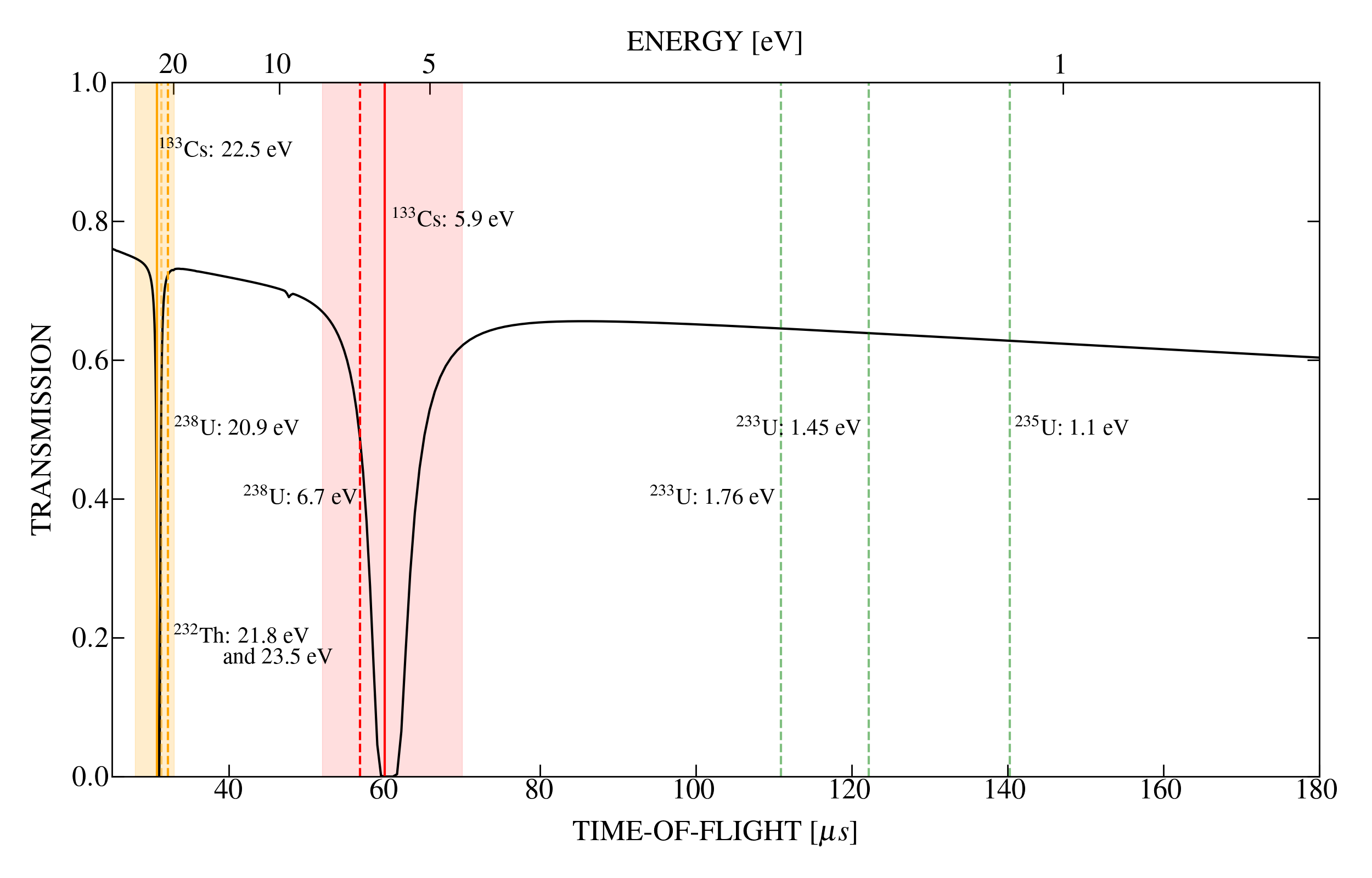}
\caption{Interference between resonances for $^{133}$Cs and different isotopes of interest for the thorium fuel cycle.}
\label{fig:clyc_lines}
\end{figure}

An additional concern presented by CLYC is the presence of resonances within the epithermal range from $^{133}$Cs in the scintillator itself. The isotope has resonances at 5.9 eV and 22.5 eV that are of particular concern, as they overlap with resonances from isotopes that are of importance to the thorium fuel cycle -- specifically, the 6.7 and 20.9 eV lines for $^{238}$U and the dual 21.8 and 23.5 eV resonances for $^{232}$Th, as seen in Fig.~\ref{fig:clyc_lines}. However, while the CLYC resonance at 5.9 eV is fully saturating, the resonance at 22.5 eV is not saturating after accounting for the effects of detector resolution and the neutron pulse -- therefore, it is possible that resonances that overlap with the latter may still be detectable, given that the target resonances will still influence a net change in counts between the open and target beams. Therefore, it is plausible that these isotopes can still be detected and quantified. 
Additionally, the ability to detect $^{233}$U and $^{235}$U, important isotopes for the thorium fuel cycle, is practically unaffected, as resonances for these isotopes are sufficiently separated from any resonances of $^{133}$Cs.

\section{Experimental Details}

Experimentation was conducted at the MIT Vault Laboratory, utilizing the NRTA setup described in~\cite{Subzwari_NRTA_Thesis_2025, mcdonald2024neutron}. The setup used a standoff distance of 2.0 m between the front face of the moderator and the front face of the detector, and the D-T generator operating at a frequency of 5 kHz, a 3.2\% duty cycle, a beam current of 40 $\mu$A, and an acceleration voltage of 110 kV. 

To characterize the CLYC detector for subsequent analysis, a background model was built for CLYC in a similar manner to the one built previously for GS20 \cite{Klein_thesis}. This involved utilizing several targets available at the Vault Laboratory with saturating resonances and extracting on- and off-axis background components, as depicted in Fig.~\ref{fig:clyc_bg}. These runs included an open beam, a 2.34 mm W target, a composite 1.50 mm In and 0.35 mm Ag target, and an off-axis run, where a sufficient amount of boron carbide was placed in the beam line in front of the detector to effectively eliminate any on-axis contributions. 

For the experimental runs, a 1.50 mm-thick tungsten (natural isotopic abundances) metal sheet was selected as the main target for analysis. In order to simulate conditions of high passive emissivity, various sources available at the Vault were considered. 

A source was desired that yielded a passive gamma rate from GS20 comparable to that from $^{233}$U in the previous measurements. Specifically, these passive gamma count rates should be comparable within the delineated ADC range corresponding to the neutron peak in PHA analysis. Furthermore, this source should produce gammas of a sufficiently wide range of energies such that they overlap in equivalent-energy with both the neutron peak in GS20 and the neutron region in CLYC -- including this stipulation allows for the possibility that gammas can contribute to the post-filter neutron signal in both cases. A $^{232}$Th source was found that satisfied these conditions, yielding a passive gamma count rate on the same order of magnitude as the previous $^{233}$U runs, and with a wide range of equivalent energies that overlapped with neutron regions for both detectors. 

As mentioned previously, tests needed to be conducted in order to determine whether pileup would present a significant issue for the CLYC detector. To do so, a 10-minute run was conducted with the detector flush against the casing for the source. The presence of significant counts in the region in PSD-ADC space corresponding to neutrons would likely imply a significant degree of unacceptable pileup. 

Regarding the setup, the $^{232}$Th source was placed in front of the boron carbide shield housing the detector, at the appropriate standoff distance needed to yield the desired count rates. A passive run was carried out with the D-T generator off, first with the GS20 detector and then with the CLYC. Subsequently, the source was removed and an open-beam run was carried out for both detectors. The 1.50 mm W target was then put in place and a target-in run was conducted, again for both detectors. Target-in measurements were conducted for 1 hour each, while open-beam measurements were 30 minutes each. 

From these measurements, four subsequent datasets were produced. One dataset for each of the detectors was simply the target-in runs, representing the expected operation of GS20 and CLYC for non-gamma-emitting targets. A second dataset for each of the detectors was also produced, where the passive run results were concatenated with those from the same target-in runs. These were used as a means of artificially reproducing a target-in-run, where the target is a significant gamma-emitter with an activity similar to the $^{233}$U samples used in previous experimentation. A combination of these datasets was used as opposed to simply running the D-T generator with both the target and $^{232}$Th simultaneously in place, as the source could not be placed sufficiently close enough to the detector to produce the desired count rate without interfering with the beam line and causing unwanted or unpredicted neutron attenuation.

\begin{figure}[p]
\captionsetup{justification=centering}
\centering
\includegraphics[width=0.9\textwidth]{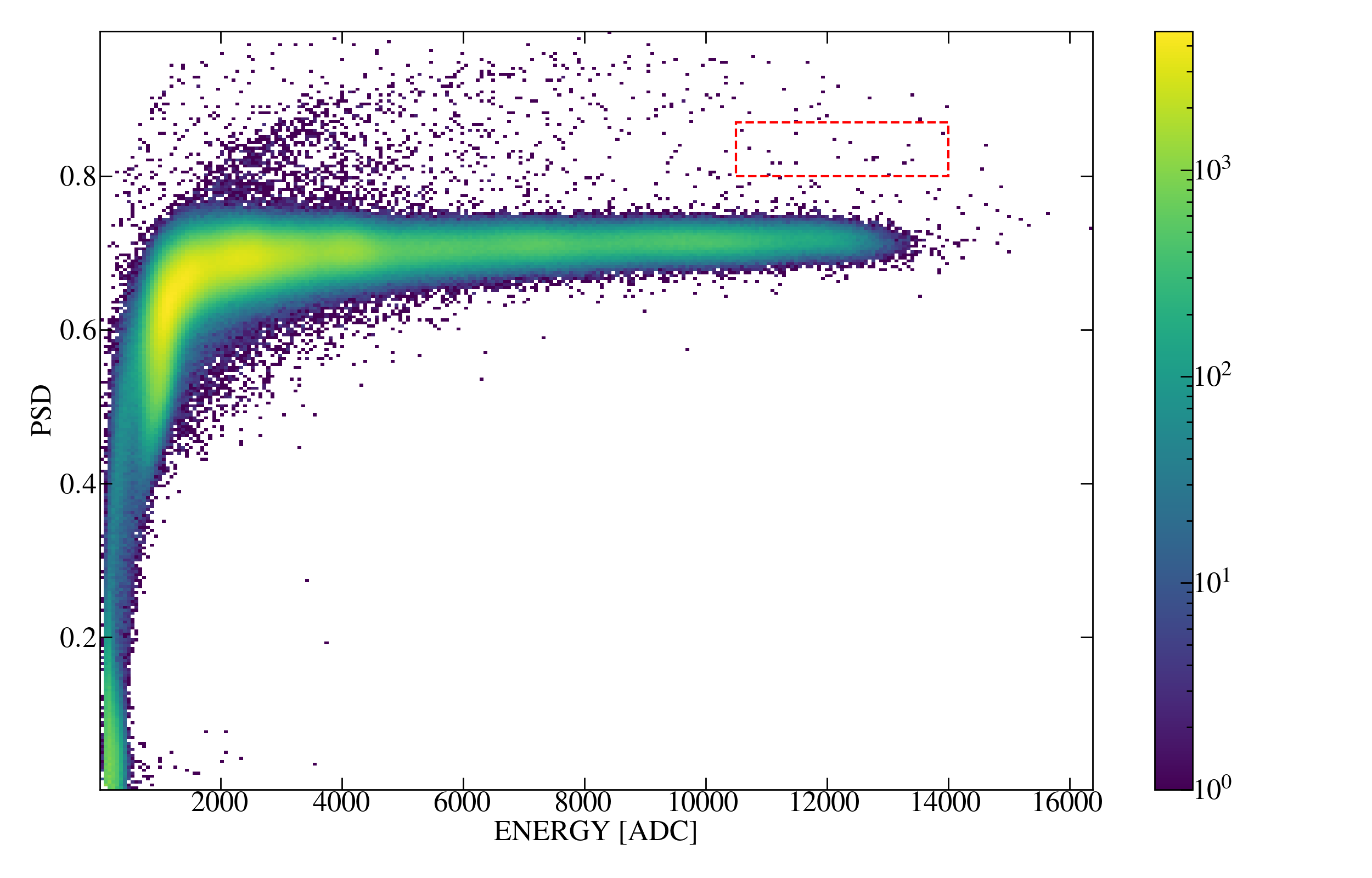}
\caption{PSD plot for the CLYC detector placed flush with the $^{232}$Th source, showing little to no pileup.}
\label{fig:no_pileup}
\end{figure}

\section{Results}

\begin{figure}[htbp]
    \centering
    \begin{subfigure}[b]{0.49\textwidth}
%\captionsetup{justification=centering}
%\centering
    \includegraphics[width=\textwidth]{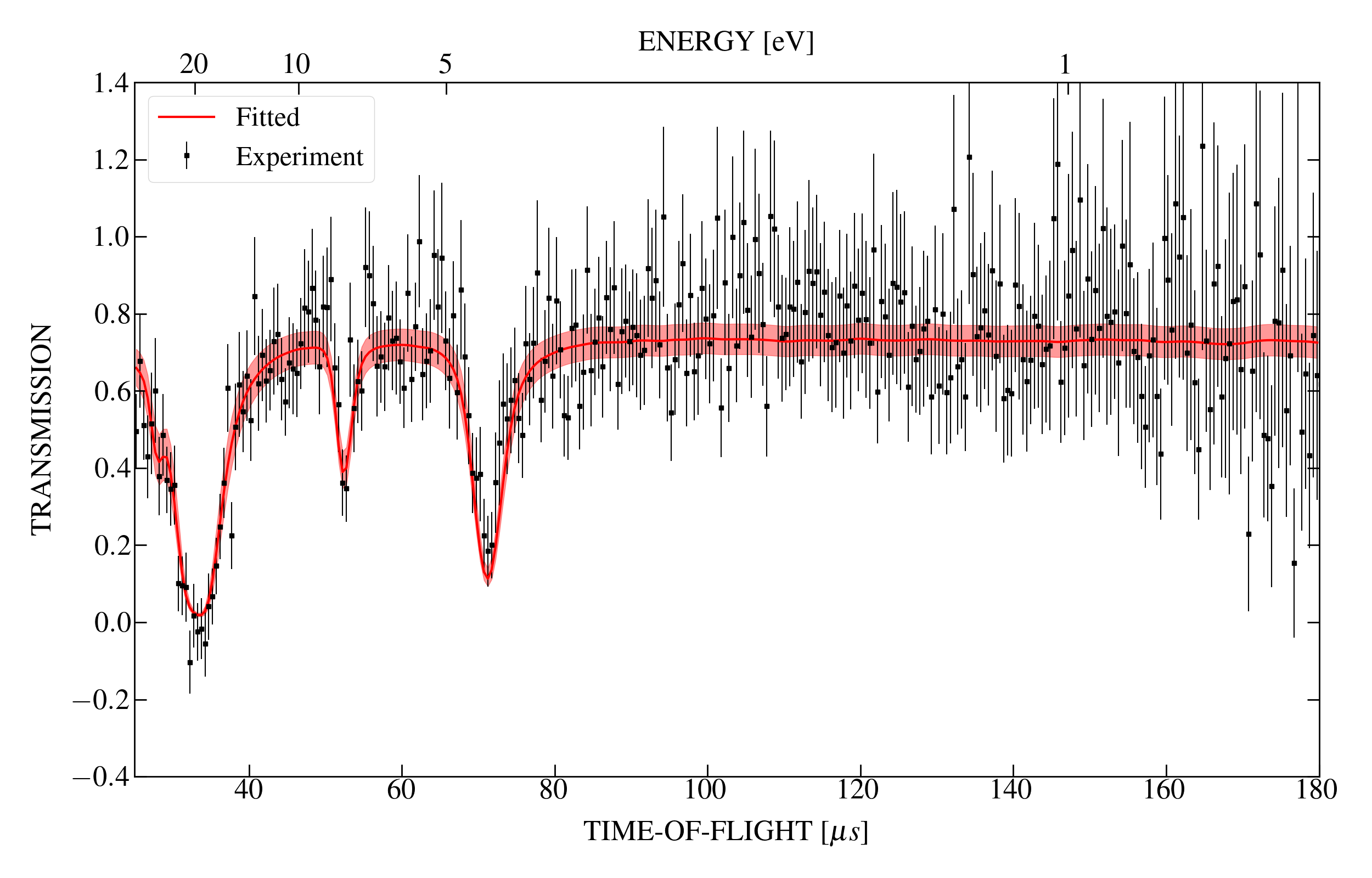}
    \caption{NeuFIT fit to a 1-hour measurement of a 1.50 mm W target using a GS20 detector.}
    \label{fig:gs20_W}
    \end{subfigure}
    \hfill
    \begin{subfigure}[b]{0.49\textwidth}
%\captionsetup{justification=centering}
%\centering
    \includegraphics[width=\textwidth]{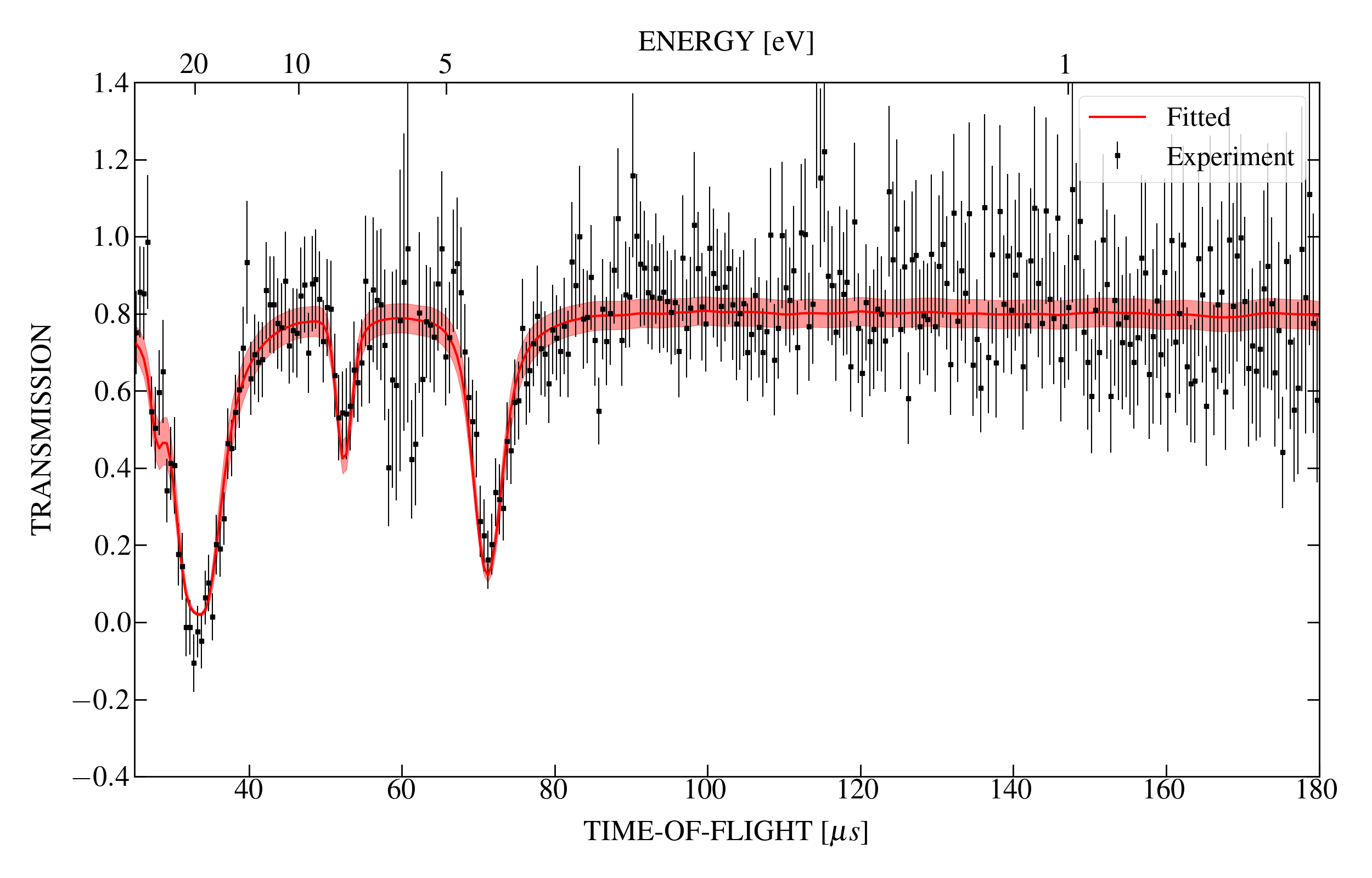}
    \caption{NeuFIT fit to a 1-hour measurement of a 1.50 mm W target using a CLYC detector.}
    \label{fig:clyc_W}
    \end{subfigure}
    \caption{Results of the 1.5 mm W target}
    \label{fig:placeholder}
\end{figure}

\begin{figure}[htbp]
    \centering
    \begin{subfigure}[b]{0.49\textwidth}
    \captionsetup{justification=centering}
    \centering
    \includegraphics[width=\textwidth]{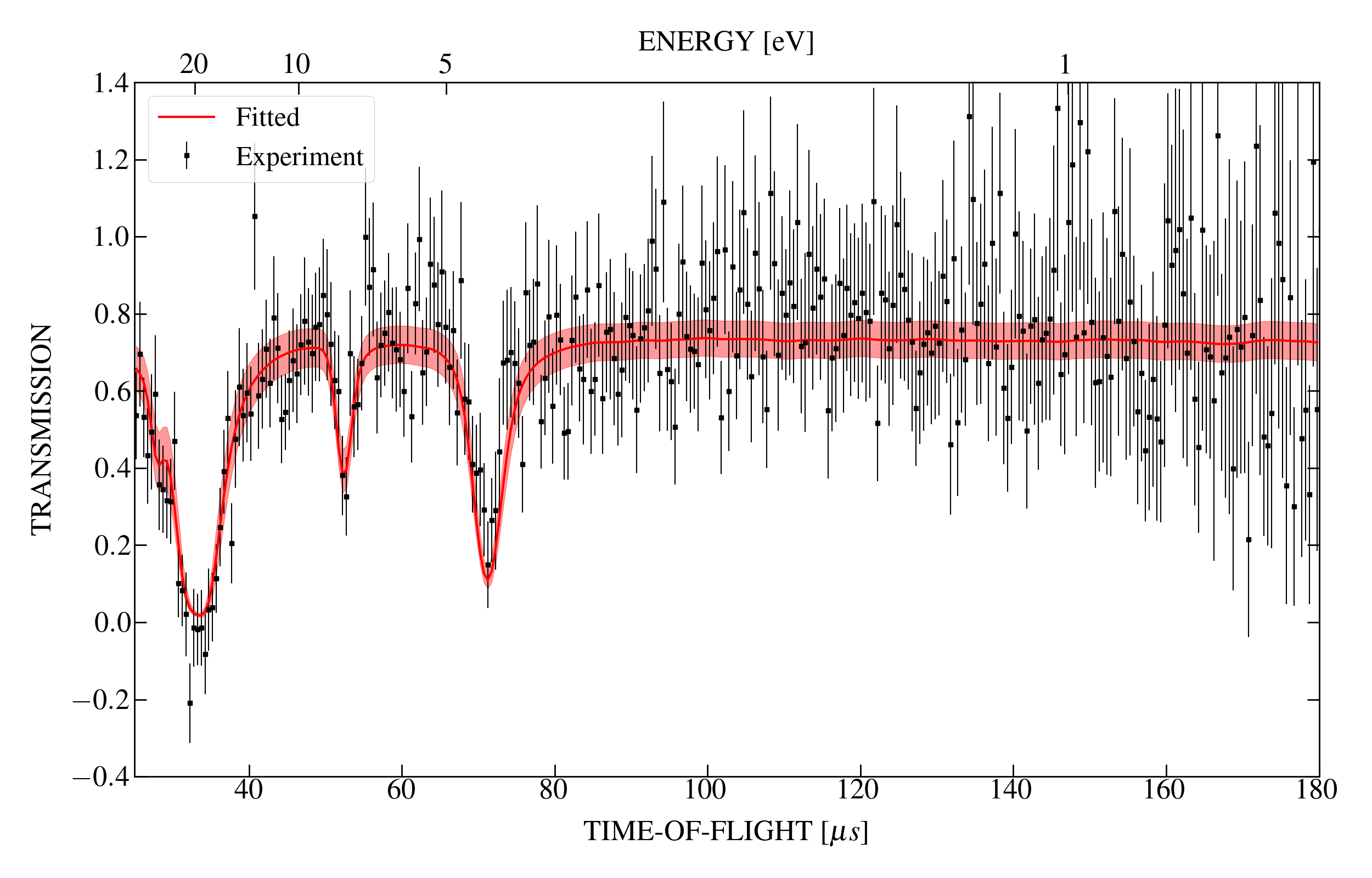}
    \caption{NeuFIT fit to a 1-hour measurement of a 1.50 mm W target with added effects of a $^{232}$Th source, using a GS20 detector.}
    \label{fig:gs20_W_Th}
    \end{subfigure}
    \hfill
    \begin{subfigure}[b]{0.49\textwidth}
    \captionsetup{justification=centering}
    \centering
    \includegraphics[width=\textwidth]{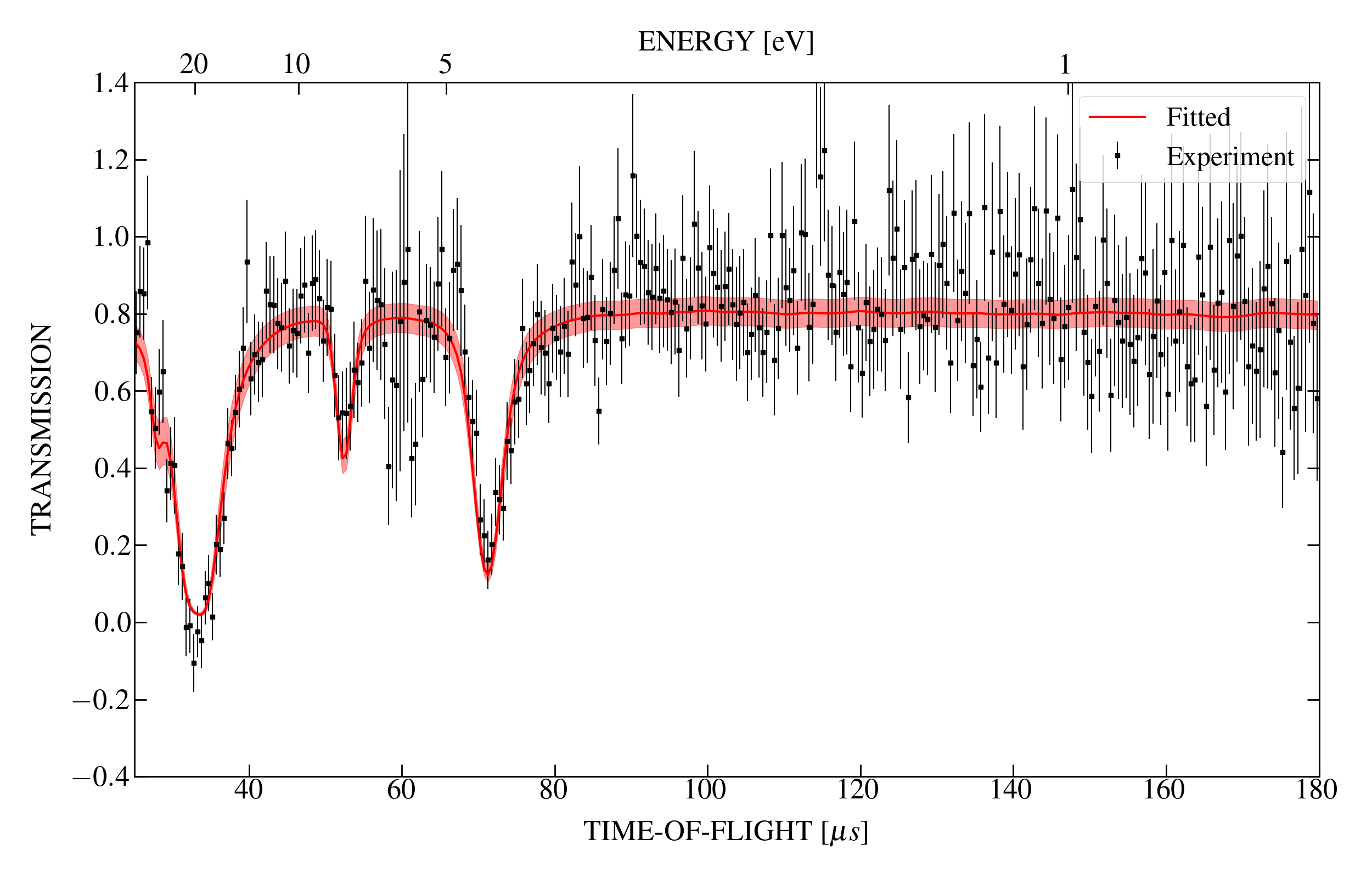}
    \caption{NeuFIT fit to a 1-hour measurement of a 1.50 mm W target with added effects of a $^{232}$Th source, using a CLYC detector.}
    \label{fig:clyc_W_Th}
    \end{subfigure}
        \caption{Results of the 1.5 mm W target with the $^{232}$Th source.}
    \label{fig:placeholder2}
\end{figure}

The previously described datasets were initially processed 
%using the Compy Python package \cite{compy}, 
utilizing the pre-existing background model for GS20 and the newly developed background for CLYC. The experimental transmission curves that were produced were analyzed using NeuFIT's Isotope Quantification Algorithm, choosing the KDE+savgol approach for transmission generation~\cite{Subzwari_NRTA_Thesis_2025, imagingreso}. The results of these fits are depicted in Figs.~\ref{fig:gs20_W} through~\ref{fig:clyc_W_Th}, with numerical results in Table~\ref{tab:transm_Th_dets}. 

In all of the plots, the tungsten resonances at 4.2, 7.6, and 19 eV can be easily seen, corresponding to approximately 70, 55, and 35 $\mu$s, respectively. The fits generally seem to adhere quite well, capturing the relative depths of each of the three resonances. Looking at the numerical results, it can be seen that all the predicted thicknesses are consistent within one standard deviation from the known value of 1.50 mm. The accuracies seem to be generally comparable. However, the uncertainties present some interesting takeaways -- when compared between CLYC and GS20, the uncertainties associated with CLYC are smaller than those from GS20, with this effect being exacerbated when the $^{232}$Th source is included. In that case, the errors are almost 50\% larger for GS20 than they are for CLYC. Furthermore, the predicted thicknesses and uncertainties for CLYC remain exactly the same given the number of significant figures -- this is conclusive of the fact that CLYC is very effective at filtering out gammas and eliminating any effect that such a gamma background would have on quantitative results. On the other hand, the uncertainty for GS20 increased significantly between when the source was absent to when it was added, demonstrating that GS20 is not as effective in filtering out the effect of gamma backgrounds. Parallel behaviors in the uncertainties can be noticed in the error bars on the corresponding figures. 

\begin{table}[ht]
\centering
\caption{Predicted thickness for a 1.50 mm W target, using GS20 and CLYC, under regular and (imitated) irradiated conditions.}
\label{tab:transm_Th_dets}
\begin{tabular}{lcc}
\toprule
& \multicolumn{2}{c}{Predicted thickness (mm)} \\
\cmidrule(lr){2-3}
Sample            & Without $^{232}$Th         & With $^{232}$Th            \\
\midrule
GS20  \quad  & $1.491 \pm 0.15\ $ & $1.560 \pm 0.19\ $ \\
CLYC  \quad  & $1.522 \pm 0.13\ $ & $1.522 \pm 0.13\ $ \\
\bottomrule
\end{tabular}
\end{table}

Furthermore, looking at Fig.~\ref{fig:no_pileup}, representing the instance where the CLYC detector was placed flush with the $^{232}$Th source, very few events are present in the PSD-ADC area corresponding to the neutron region. While these events may be cosmic neutrons, pileup, or other events, it is indicative of the fact that pileup, presented by an excess of counts, is not a significant source of concern for the given setup and expected count rates.

\section{Conclusion}

Experimental results have demonstrated that GS20 and CLYC are effective detectors for a portable NRTA setup, although they offer varying advantages and disadvantages. While GS20 is useful for its efficiency and fast response times, it has high gamma sensitivity and limited neutron-gamma discrimination. On the other hand, CLYC requires greater thicknesses to obtain efficiencies on par with GS20, yet has extremely effective neutron-gamma discrimination, making for an acceptable tradeoff. Additionally, while CLYC is a slower detector overall, it has been determined that issues of pileup at the count rates seen in this particular NRTA setup and for gamma-emitting targets of interest are not a cause for concern. 

Comparing the operation of GS20 and CLYC, both detectors are able to accurately quantify the amount of tungsten in a given target, for both a high gamma-emitting target and a non-emitting one. However, CLYC can do so more precisely, with a large difference in the relative precision when the target is gamma-emitting -- this presents the conclusion that CLYC is likely more appropriate for these particular applications. As described in~\cite{Subzwari_NRTA_Thesis_2025}, if additional composite targets are utilized, some of which may have large passive gamma signatures, the precision on the predicted thicknesses may increase significantly, frustrating the ability to confidently quantify and identify the presence of certain isotopes. In such cases, reducing the uncertainties on the predicted values may be critical, presenting CLYC as a promising option.

A preliminary consideration was conducted regarding the resonances present from $^{133}$Cs within CLYC and potential interference with resonances from isotopes of interest for advanced reactor safeguards. However, further experimentation is likely needed to confirm that accurate fitting can still be performed, especially at the non-saturating 22.5 eV line. Such experimentation will bolster the applicability of CLYC for such contexts.

There are a number of factors that need to be considered regarding the choice of an optimal detector for NRTA. While GS20 has been used previously for this portable NRTA setup, the use of CLYC for NRTA is a novel approach. These results demonstrate that CLYC can provide distinct advantages, especially in the case of advanced reactor safeguards and high gamma-emitting environments. Further testing and experimentation will serve to reinforce and advance the feasibility of NRTA as a whole, both within these specialized considerations and in broader applications.  Future work should also compare the recently developed LiI:Ce scintillators, as they have a high neutron capture cross section while achieving PSD capability. 

\section*{Acknowledgments}
This research was supported by the U.S. National Nuclear Security Administration (NNSA) Office of Defense Nuclear Nonproliferation Research and Development within the U.S. Department of Energy (DOE) under Contract DE-AC05-76RL01830 and PNNL-SA-221659. This work was also supported at MIT by NNSA through the Monitoring Technology and Verification (MTV) consortium, award number DE-NA0003920.  

\bibliography{refs}

\end{document}